\documentclass[12pt,a4paper,reqno]{amsart}
\usepackage{xypic,a4wide,amsmath,amssymb,amsthm}
\usepackage[utf8x]{inputenc}

\usepackage[T1]{fontenc}

\usepackage{amssymb}
\usepackage{amsmath}

\usepackage[dvips]{graphicx}
\usepackage{array}

\title[\ ]{Darboux solutions  of non-abelian 
quantum Painlev\'e II equation in terms of quasideterminants}

\author{Irfan Mahmood}
\address{
CHEP, University of the Punjab, 54590 Lahore,  Pakistan} \email{mahirfan@yahoo.com} \urladdr{}

\begin{document}

\maketitle

\begin{abstract}
In this article non-abelian version of  quantum Painlev\'e II equation is presented with  Its quasideterminant solutions  has been  derived by using the Darboux transformations. This non-abelian  quantum Painlev\'e II equation may be considered as  a specific case of  its purely  noncommutatie analogue presented by V. Retakh and V. Rubtsov .   In these computations the  quantum Painlev\'e II symmetric form  with commutation  relations  presented by  H. Nagoya are applied to derive  Nonabelian quantum Painlev\'e II equation and a new commutation relation between   variable  $z$ and the solution   $ f(z)$  such as  $ z f - f z = \frac{1}{2} i \hbar f $   is presented. Finally,  the  Darboux solutions of that system  are generalized to the $N$-th form in terms of quasideterminants.
\end{abstract}
\section{Introduction}

The Painlev\'e equations have got considerable attention because of their wide applications in several areas of mathematics and physics. For example, an important fact was observed in \cite{ reb, rdg}  concerning the application of PI equation  in two dimensional quantum gravity where its free energy functional can be expressed in terms of PI solution.  Further it was observed by V. E. Zakharov, E. A. Kuznetsov and S. L. Musher  \cite{rasf}  that  the PII equation is helpful  to describe the  strong wave collapse in the  case of  three-dimensional nonlinear Schr\"odinger equation and one of its interesting applications has been studied  by A. S. Fokas and S. Tanveer \cite{rasf}  in the analysis of classical  Hele-Shaw problems.  More over it has been studied that the solutions of some field theoretical problems  can be obtained in terms of Painlev\'e transcendents for example,  these transcendents appear in the  description of    two-point correlation function for the two dimensional Ising model as presented in  \cite{rebm} and also useful in the analysis of  two-point correlation function at zero temperature for the case of one dimensional Bose-gas, details can be found in  \cite{rmj}. The more recent and interesting applications of the modern theory of  Painlev\'e  equations include differential geometry of surfaces, orthogonal polynomials and string theory , see  \cite{ rasf, ras, rri} fro detail. The classical 
 Painle\'e equations are also  regarded as completely integrable equations as mentioned by  \cite{r5, r11, r12}  that they admit  linear representations, possess Hamiltonian structures. and obeyed the Painlev\'e test.

A very interesting aspects of these equations is that they  also arise as ordinary differential equations (ODEs)   reductions of some integrable systems, i.e, N. Joshi and M Mazzocco \cite{r4, MA}  has been shown the the ODE reduction of the KdV equation is Painlev\'e II (PII) equation.

The  quantum analogue of the classical Painlev\'e II equation  has been studied in  \cite{NH} where  the  Painlev\'e II  symmetric form  
\begin{equation}\label{eq:1} 
\left\{
\begin{array}{lr}
 f^{'}_{0}= f_{0}f_{2} + f_{2}f_{0} + \alpha_{0} \\
 f^{'}_{1}= -f_{1}f_{2} - f_{2}f_{1} + \alpha_{1}\\
f^{'}_{2}= f_{1} - f_{0}
\end{array}
\right.
 \end{equation} 
 which  possesses the affine  Weyl group symmetry of type $ A^{1}_{l}$  applied for the quantization of   classical Painlev\'e II (PII) equation.  Further after having the quantization of  the classical Painlev\'e equations the extension of these equations to the  noncommutative (NC) spaces got considerable attention
 A  Very  initial achievement in NC direction was obtained by V. Retakh and V. Roubtsov in \cite{7} where they have introduced purely NC version of PII equation 
 
 \begin{equation}\label{VRVR} 
  f^{''}_{2} = 2 f^{3}_{2}- 2[z,f_{2}]_{+} + 4 ( \beta + \frac{1}{2} ) 
 \end{equation}

  with the help of its symmetric form in their  case the fields $ f_{0} $, $ f_{1} $,  $ f_{2} $  obey a kind of star product, where these fields are  purely non-commutative.   Further, the quasideterminant solutions of that NC PII equation  have been  constructed by M. Irfan \cite{MIRFAN}  through its zero-curvature condition and later on these solutions are calculated in \cite{IM} by  taking NC Toda soutions as seed in NC PII  Darboux transformations .\\ 
In this article I have  extended the H. Nagoya  \cite{NH}  work on quantum PII equation to derive its non-abelian analogue. 
Here I have introduced a zero-curvature condition  that is equivalent to the non-abelian quantum PII equation
\begin{equation}\label{L8} 
\left\{
\begin{array}{lr}
 f^{''}_{2} = 2 f^{3}_{2}- 2[z,f_{2}]_{+} +  c\\
 z f_{2} - f_{2} z = \frac{1}{2} i \hbar f_{2}
\end{array}
\right.
 \end{equation} 
these derivations are also involved  the symmetric form (\ref{eq:1})  of PII equation and  the fields  $ f_{0} $, $ f_{1} $,  $ f_{2} $  obey the quantum commutation relation  given in  \cite{NH} .  The basic difference between the quantum  PiI equation   and  the non-abelian quantumPII equation  (\ref{L8})  is that here the  variable $z$ and field $f_{2}$ appear as  noncommuting  elements but in cacs of H. Nagoya \cite{NH}  these elements treated classically, as  commuting variables. Further this can be shown  that under the classical limit   $\hbar  \rightarrow 0$ the system   (\ref{L8})  reduces to its classical analogue. I have also constructed it Riccati form and Further  its Darboux solutions have also be derived. Finally these solutions are generalized to the $N$-th form in terms of quasideterminants.  

\section{Zero curvature representation of non-abelian quantum PII equation}
 In the theory of Integrable systems, we are  familiar with the Lax formalism , first time was introduced in \cite{PL} and zero curvature representations of these systems. Both the  Lax technique  and zero curvature condition  are extensively applied to construct the solutions of Integrable systems in classical as well as in NC case. These representations involve two linear operators, these operators may be differential operators or matrices. The application of these formalisms to explore the  various integrable aspects of nonlinear systems has been studied in wide spectrum, see, for example  \cite{r22, r19, r16, r13,  r24, r23}.
If $A$ and $B$ are the linear operators  and subjected to a linear system   $A(x,t)\Psi = \lambda \Psi_{t} $,  $\Psi_{t}=B(x,t)\Psi$   then Lax equation is given by  $ A_{t} = [B,A]$  where $ \lambda$ is spectral parameter and $[B,A]$ is commutator. The compatibility condition of inverse scattering problem  $\Psi_{x}=A(x,t)\Psi$ and $\Psi_{t}=B(x,t)\Psi$ yields $ A_{t}-B_{x}= [B,A]$  which is called the zero curvature representation of integrable systems , which has been applied to many classical and NC systems, for a brief description see  \cite{e2, 4, 3,1}. In this article the commutator  and anti-commutator will be written as  $[,]_{-}$ and  $[,]_{+}$ respectively. Now the Lax equation and zero curvature condition can be expressed as  $ A_{t} = [B,A]_{-}$ and  $ A_{t}-B_{x}= [B,A]_{-}$.  \\
\textbf{Proposition 1.1.}\\
The compatibility condition of following linear system
\begin{equation}\label{QPIIb} 
 \Psi_{\lambda}=A(z;\lambda)\Psi , \; \; \;  \Psi_{z}=B(z;\lambda)\Psi 
\end{equation}
with Lax matrices
\begin{equation}\label{RDTa}
\left\{
\begin{array}{lr}
A= (8i \lambda^{2} + if^{2}_{2} - 2iz ) \sigma_{3} + f^{'}_{2} \sigma_{2}+  (\frac{1}{4} c \lambda^{-1} -4\lambda f_{2} )\sigma_{1} + i \hbar \sigma_{2} \\
B =  -2i \lambda \sigma_{3}  + f_{2} \sigma_{1}  + f_{2}I
\end{array}
\right.
 \end{equation} 
yields non-abelian quantum PII equation with quantum commutation relations  given in \cite{NH}  , here $I$ is $ 2 \times 2$ identity matrix and $ \lambda $ is spectral parameter and $c$ is constant.\\
\textbf{Proof:}\\
The compatibility condition of system (\ref{QPIIb}) yields zero curvature condition
\begin{equation}\label{ZC4} 
 A_{z}-B_{\lambda}= [B,A]_{-}.
\end{equation} 
 We can easily  evaluate the values for  $ A_{z}$, $B_{\lambda}$ and $[B,A]_{-} = BA -AB$ from the linear system (\ref{RDTa}) as follow
\begin{equation}\label{V1} 
 A_{z} = ( i f^{'}_{2} f_{2} + if_{2} f^{'}_{2}  -2i )\sigma_{3} + f^{''}_{2} \sigma_{2} - 4 \lambda f^{'}_{2} \sigma_{1} 
\end{equation} 

\begin{equation}\label{V2} 
 B_{\lambda} =  -2i \sigma_{3}  
\end{equation} 
and

\begin{equation}\label{V3} 
[ B,A ]=   \begin{pmatrix}
 i f^{'}_{2} f_{2} + if_{2} f^{'}_{2} + [f_{2},z]_{-} - i  \hbar  &  \delta^{+}  \\
  \delta^{-}  & - i f^{'}_{2} f_{2} - if_{2} f^{'}_{2} [z, f_{2}]_{-} + i  \hbar 
\end{pmatrix}
\end{equation} 

where
\[  \delta^{+} =  -i f^{''}_{2}+2if_{2}^{3}-2i[z,f_{2}]_{+} +ic  + i [ f^{'}_{2}, f_{2}]_{-} + 4i\lambda  \hbar  \]

and

\[   \delta^{-} = i f^{''}_{2}- 2if_{2}^{3}+2i[z,f_{2}]_{+} - ic  + i [f_{2}, f^{'}_{2} ]_{-}- 4i\lambda   \hbar.\]
now after substituting these values from (\ref{V1}), (\ref{V2}) and (\ref{V3}) in equation  (\ref{ZC4}) we get

\begin{equation}\label{RM1} 
 \begin{pmatrix}
[ f_{2},z]_{-} -\frac{1}{2} i  \hbar  &   \delta^{+}   \\
  \delta^{-}  & [z, f_{2}]_{-} + \frac{1}{2} i  \hbar 
\end{pmatrix}=0
\end{equation} 
and the above result (\ref{RM1}) yields the following expressions   
\begin{equation}\label{L4} 
 [ f_{2},z] = \frac{1}{2}i  \hbar f_{2} 
\end{equation} 
and 
\begin{equation}\label{L5} 
 i f^{''}_{2}- 2if_{2}^{3} + 2i[z,f_{2}]_{+} - ic  + i [f_{2}, f^{'}_{2} ]_{-} - 4i\lambda \hbar=0
\end{equation} 
equation (\ref{L4}) shows quantum relation between the variables $ z$ and $ f_{2}$. In equation (\ref{L5}) the term $ i [f_{2}, f^{'}_{2}]_{-} - 2i\lambda \hbar $ 
can be  eliminated  by using  equation $ f^{'}_{2} = f_{1} - f_{0}$ from  (\ref{eq:1} )   and  quantum commutation relations in \cite{NGR} . For this purpose let us replace 
$ f_{2}  $ by $ -  \frac{1}{2}\lambda^{-1}  f_{2}$ ,  then commutation relations become 
\begin{equation}\label{L6} 
[f_{0},f_{2}]_{-}=[f_{2},f_{1}]_{-}=  -2 \lambda \hbar.
\end{equation}
Now let us take the commutator of the both side of the equation  $ f^{'}_{2} = f_{1} - f_{0}$  with $ f_{2}$ from right  side, we get

\[ [ f^{'}_{2}, f_{2} ]_{-} = [ f_{1}, f_{2}]_{-} - [ f_{0} , f_{2}]_{-} \]
above equation  with the commutation relations (\ref{L6}) can be written as
\begin{equation}\label{L7} 
  [ f^{'}_{2}, f_{2} ]_{-}= - 4 \lambda \hbar .
\end{equation}
Now after substituting the value of $   [ f^{'}_{2}, f_{2} ]_{-} $ from (\ref{L7}) in (\ref{L5}) we get 
\[  i f^{''}_{2}- 2if_{2}^{3} + 2i[z,f_{2}]_{+} - ic =0.\]
Finally, we can say that  the compatibility of condition of linear system (\ref{QPIIb}) yields the following expressions
\begin{equation}\label{L8} 
\left\{
\begin{array}{lr}
 f^{''}_{2} = 2 f^{3}_{2}- 2[z,f_{2}]_{+} + c \\
 z f_{2} - f_{2} z = \frac{1}{2} i \hbar f_{2}
\end{array}
\right.
 \end{equation} 
in above system (\ref{L8}) the first equation can be considered as nonabelian version of quantum Painlev\'e II equation that is equipped with  a quantum commutation relation and that 
equation can be  reduced to the classical PII equation under the classical  limit when $  \hbar \rightarrow 0$.\\
\textbf{Remark 1.2.}\\
The linear system (\ref{QPIIb}) with eigenvector $ \Psi = \begin{pmatrix}
\chi  \\
 \phi
\end{pmatrix}$ and setting $ \varDelta = \chi \psi^{-1}_{2}$ can be reduced to the PII Riccati equation associated to the nonabelian quantum PII equation
 \[  \varDelta^{'} = -4i \varDelta + f_{2}  + [ f_{2},\varDelta ]_{-}  - \varDelta f_{2}  \varDelta .\]
\textbf{Proof:}\\
In order to derive the Riccati equation associated to the system (\ref{L8}) ,we apply   the method of Konno and Wadati \cite{KKMW}. For this purpose let us substitute the eigenvector 
$ \Psi = 
\begin{pmatrix}
\chi  \\
 \phi
\end{pmatrix} $ in linear systems (\ref{QPIIb}) and  we get
\begin{equation} \label{NCQPIIa}  
 \left \{\begin{aligned}
\frac{d \chi  } {d\lambda} = ( 8  i \lambda^{2} +  i f^{2}_{2}-2i z)  \chi+ ( - i f^{'}_{2}+\frac{1}{4} C_{0} \lambda^{-1}-4 \lambda f_{2}  + \hbar ) \phi\\
\frac{d\phi}{d\lambda} = (i f^{'}_{2} +\frac{1}{4} C_{0} \lambda^{-1}-4 \lambda f_{2} - \hbar)  \chi+ (-8  i\lambda^{2} - i f^{2}_{2} +2i z)\phi
\end{aligned}
 \right. 
 \end{equation} 
 and  
 \begin{equation}\label{NCQPIIb}   \left \{ \begin{aligned} \chi^{'} =( -2i \lambda + f_{2} ) \chi + f_{2} \phi \\ \phi^{'} =f_{2} \chi + ( 2i \lambda  + f_{2} ) \phi \end{aligned} \right.  \end{equation} 
 where $ \chi^{'} = \frac{d \chi } {dz} $ and  now from system (\ref{NCQPIIb}) we can derive the following expressions  
 \begin{equation}\label{NCQPIIc} 
\chi^{'} \phi^{-1} =  (-2i \lambda  + f_{2}  ) \chi \phi^{-1}   +  f_{2}   
 \end{equation} 
 \begin{equation}\label{NCQPIId} 
 \phi^{'} \phi^{-1} =  -2i \lambda  + f_{2}  + f_{2}  \chi \phi^{-1}.
 \end{equation}
 Let consider the following substitution \begin{equation}\label{NCQPIIe}  \varDelta = \chi \phi^{-1} \end{equation} now taking the derivation of above equation with respect to $z$\[\varDelta^{'} = \chi ^{'} \phi^{-1} -  \chi\phi^{-1} \phi^{'} \phi^{-1}\] after using the ( \ref{NCQPIIc}), (\ref{NCQPIId}) and (\ref{NCQPIIe}) in above equation we obtain\begin{equation}\label{NCQPIIf}  \varDelta^{'} = -4i \varDelta + f_{2}  + [ f_{2},\varDelta ]_{-}  - \varDelta f_{2}  \varDelta \end{equation} the  above expression (\ref{NCQPIIf}) can be regarded as quantum Riccati equation in $ \varDelta$ related to the nonabelian quantum PII equation which has been derived from its associated linear systems (\ref{NCQPIIb}). In next section the Darboux solutions to no-abelian quantum PII equation are constructed by using the idea of  Darboux transformations presented in  \cite {VM}.
 \\\section{Darboux transformation for nonabelian quantum PII equation}
\textbf{Proposition 2.1.}\\
The Darboux transformation for the solution $ u$ of non-abelian quantum PII equation  (\ref{L8})  with the help of its associated  linear system can be constructed in the following form
\[ u[1]= - 4\lambda \Phi_{1} \chi ^{-1} +  \Phi_{1} \chi_{1}^{-1} u \Phi_{1} \chi^{-1}_{1} \]
here $ u[1]  $ is a new solution of QP-II equation generated by initial solution $u$, here  $f_{2}$ has been replaced by  $ u $, just for a simple notation.\\
\textbf{Proof:}\\
For the derivation of  non-abelian QP-II  Darboux transformation we consider the linear system  (\ref{QPIIb}) with a column vector $\psi=
\begin{pmatrix}
\chi  \\
 \Phi
\end{pmatrix}$.
Now the linear system will become
\begin{equation}\label{di}
 \begin{pmatrix}
\chi  \\
 \Phi
\end{pmatrix}_{\lambda} =\begin{pmatrix}
8  i \lambda^{2} +  i  u^{2} -2i z & -i  u _{z}+\frac{1}{4}C \lambda^{-1}-4\lambda  u + \hbar  \\
 i  u_{z}+\frac{1}{4}c  \lambda^{-1} - 4\lambda  u - \hbar & - 8  i\lambda^{2} -  i  u^{2}+2i z 
\end{pmatrix}
\begin{pmatrix}
 \chi  \\
 \Phi
\end{pmatrix}
\end{equation}
\begin{equation}\label{d2}
 \begin{pmatrix}
 \chi  \\
\Phi
\end{pmatrix}_{z} =
\begin{pmatrix}
 -2i \lambda +u & u \\
u & 2i \lambda + u
\end{pmatrix}
\begin{pmatrix}
 \chi  \\
 \Phi
\end{pmatrix}.
\end{equation} The standard transformations on $ \chi$ and $\Phi$ are given below
\begin{equation}\label{d3}
\chi \rightarrow \chi[1]= \lambda \Phi-\lambda_{1}\Phi_{1}(\lambda_{1}) \chi_{1}^{-1}(\lambda_{1}) \chi
\end{equation}
\begin{equation}\label{d4}
\Phi \rightarrow \Phi[1]=\lambda \chi-\lambda_{1} \chi_{1}(\lambda_{1})\Phi_{1}^{-1}(\lambda_{1})\Phi
\end{equation}
where $ \chi $ , $\Phi $ are arbitrary solutions at  $\lambda$ and  $ \chi_{1}(\lambda_{1})$ ,  $\Phi_{1}(\lambda_{1})$ are the particular solutions at $\lambda=\lambda_{1}$ of 
equations (\ref{di}) and (\ref{d2}), these equations will take the following forms under the transformations (\ref{d3}) and (\ref{d4})
\begin{equation}\label{d5}
 \begin{pmatrix}
  \chi[1]  \\
 \Phi[1]
 \end{pmatrix}_{\lambda} =
\begin{pmatrix}
 8i\lambda^{2} + i u^{2}[1]-2i z & b_{+}  \\
b_{-} &  -8 i\lambda^{2} - i u^{2}[1]+2i z 
\end{pmatrix}
\begin{pmatrix}
 \chi[1]  \\
 \Phi[1]
\end{pmatrix}
\end{equation}
where
\[ b_{+} = -i u_{z}[1]+\frac{1}{4}C \lambda^{-1}-4\lambda u[1]  +  \hbar\]
 and
 \[ b_{-} = iu_{z}[1]+\frac{1}{4}C \lambda^{-1}-4\lambda u[1]  +  \hbar\]
\begin{equation}\label{d6}
 \begin{pmatrix}
  \chi[1]  \\
 \Phi [1]
 \end{pmatrix}_{z} =
\begin{pmatrix}
 -2i \lambda +u[1] & u[1] \\
u[1] & 2i \lambda +u[1]
\end{pmatrix}
\begin{pmatrix}
 \chi[1]  \\
 \Phi[1]
\end{pmatrix}.
\end{equation}
Now from  (\ref{d2}) and equation (\ref{d6}) we have the following expressions
\begin{equation}\label{d7}
\chi_{z}=(-2 i \lambda  +  u )\chi + u \Phi
\end{equation}
\begin{equation}\label{d8}
\Phi_{z}=(i\lambda  + u )\Phi +  u \chi
\end{equation}
and
\begin{equation}\label{d9}
\chi_{z}[1]=-(2i\lambda + u[1] )\chi[1] + u[1] \Phi[1]
\end{equation}
\begin{equation}\label{d10}
\Phi_{z}[1]=(2i\lambda  + u[1] ) \Phi[1]+u[1] \chi[1].
\end{equation}
Now  substituting the transformed values  $ \chi[1] $ and  $\Phi[1] $  in equation (\ref{d9}) and then after using the  (\ref{d7}) and (\ref{d8}) in resulting equation, we get
\begin{equation}\label{d11}
u[1]= - 4 \lambda \Phi_{1} \chi^{-1}_{1}  +  \Phi_{1} \chi_{1}^{-1} u \Phi_{1} \chi^{-1}_{1} .
\end{equation}
Equation (\ref{d11}) represents the Darboux transformation of QPII equation, where $v[1]$ is a new solution of QPII equation, this shows that how the new solution is 
related to the seed solution $u$. By applying the DT iteratively we can construct the multi-soliton solution of QPII equation.

\section{A Brief Introduction of Quasideterminants}
This section is devoted to a brief review of quasideterminants introduced in  \cite{GelRet}. Quasideterminants are the replacement to classical determinant of the matrices with noncommutative entries and these determinants play very important role to construct the multi-soliton solutions of NC integrable systems such as for the case of NC KP equation these determinats are applied by C. R. Gilson, J. J. C. Nimmo  \cite{r26}  to construct its solutions. Quasideterminants are not just a noncommutative generalization of usual commutative determinants but rather related to inverse matrices, quasideterminants for the square matrices are defined as
if $A = a_{ij}$ be a $ n \times n $ matrix and $B = b_{ij}$ be the inverse matrix of A. Here all
matrix elements are supposed to belong to a NC ring with an associative
product.
Quasideterminants of $A$ are defined formally as the inverse of the elements of $B = A^{-1}$
\[ |A|_{ij}=b^{-1}_{ij} \] this expression under the limit $ \theta^{\mu \nu} \rightarrow 0 $ , means entries of $A$ are commuting, will reduce to \[ |A|_{ij}= (-1)^{i+j}\frac{detA}{detA^{ij}} \]
where $A^{ij}$ is the matrix obtained from $ A $ by eliminating  the $ i$-th row and the $ j$-th column.
We can write down more explicit form of quasideterminants. In order to see it, let us
recall the following formula for a square matrix
\begin{equation} A =  \left(\begin{array}{cc} A  & B   \\ C  & D   \end{array}\right)^{-1}
=  \left(\begin{array}{cc}  A-BD^{-1}C)^{-1} & -A^{-1}B(D-CA^{-1}B)^{-1}  \\ -(D-CA^{-1}B)^{-1} CA^{-1} & (D-CA^{-1}B)^{-1}   \end{array}\right) \end{equation}
where $A$ and $D$ are square matrices, and all inverses are supposed to exist. We note that
any matrix can be decomposed as a $
2\times 2$ matrix by block decomposition where the diagonal
parts are square matrices, and the above formula can be applied to the decomposed $2 \times 2$
matrix. So the explicit forms of quasideterminants are given iteratively by the following
formula
\[ \vert A \vert _{ij}=a_{ij}-\Sigma_{p\neq i , q\neq j}  a_{iq} \vert {A}^{ij} \vert^{-1} _{pq} a_{pj}  \]
the  number of quasideterminant of a given matrix will be equal to the numbers of its elements for example a matrix of order $3$ has nine quasideterminants. It is sometimes convenient to represent the quasi-determinant as follows
\begin{equation} \label{QDD} 
 \vert A \vert_{ij}=  \left \vert   \begin{array}{ccccc} a_{11} &  \cdots  &  a_{1j}   & \cdots  & a_{1n}   \\ 
 \vdots & \vdots   & \vdots    & \vdots  & \vdots \\ 
 a_{i1} & \cdots   & \fbox{$ a_{ij} $}  & \cdots  & a_{in} \\  
  a_{in} & \cdots   & a_{ni}      & \cdots  & a_{nn}   \end{array} \right \vert . \end{equation}
Let us consider  examples of matrices with order $2$ and $3$, for $2\times2$ matrix
\[ A = \left(\begin{array}{cc}  a_{11} & a_{12}   \\           a_{21} & a_{22}  \end{array}\right) \]
now the quasideterminats of this matrix are given below
\[  \vert A \vert _{11}=  \left \vert  \begin{array}{cc}  \fbox{$a_{11}$}  & a_{12}  \\  a_{21} & a_{22}  \end{array}\right \vert   = a_{11} - a_{12} a^{-1}_{22} a_{21} \]

 \[  \vert A \vert _{12}=  \left \vert  \begin{array}{cc}  a_{11} &  \fbox{$a_{12}$} \\  a_{21} & a_{22}  \end{array}\right \vert = a_{12} - a_{22} a^{-1}_{21} a_{12} \]

\[  \vert A \vert _{21}=  \left \vert  \begin{array}{cc} a_{11} & a_{12} \\    \fbox{$a_{21}$} & a_{22}  \end{array}\right \vert =  a_{21} - a_{11} a^{-1}_{12} a_{22}  \]

 \[  \vert A \vert _{22}=  \left \vert  \begin{array}{cc} a_{11} & a_{12} \\   a_{21}& {\boxed{a_{22}}}  \end{array}\right \vert = a_{22} - a_{21} a^{-1}_{11} a_{12}. \]
 The number of quasideterminant of a given matrix will be equal to the numbers of its elements for example a matrix of order $3$ has nine quasideterminants.
Now we consider the example of $3\times3$ matrix, its first quasidetermints can be evaluated in the following way\\

$ \vert A \vert _{11}=  \left \vert  \begin{array}{ccc} {\boxed{a_{11}}} & a_{12} & a_{13} \\   a_{21} & a_{22} & a_{23}\\  a_{31} & a_{32} & a_{33}  \end{array}\right \vert  
= a_{11}-a_{12} M a_{21}-a_{13} M a_{21}-a_{12} M a_{31}-a_{13} M a_{31}$\\
where $ M =   \left \vert  \begin{array}{cc} {\boxed{a_{22}}} &  a_{23}\\   a_{32} & a_{33}\end{array}\right \vert^{-1} ,$ similarly we can evaluate the other eight quasideterminants of this matrix.\\  
\\
\section{ Quasideterminant representation of Darboux transformation}
\textbf{Proposition 7.4.}\\
The $N$-fold Darboux transformation for non-abelian QPII solution can be derived by the iteration of (\ref{d11} )in the following form, taking $ \lambda = \gamma $
\begin{equation}\label{DT1} 
 u[N+1]=-4\gamma \Pi^{N}_{k=1} \Theta_{k}[k]+ \Pi^{N}_{k=1} \Theta_{k}[k] u[1] \Pi^{1}_{j=N}\Theta_{j}[j] \;\;\;\; \text{for} \;\; N \geq 1
\end{equation} 
with 
\[\Theta_{N}[N]=\gamma_{N}^{\phi}[N] \gamma_{N}^{\chi}[N]^{-1}\]
where $u[1]$ is seed solution and $  u[N+1]$ are the new solutions of QPII equation \cite{MIRFAN} and $ \gamma_{N}^{\phi}[N] $ ,  $ \gamma_{N}^{\chi}[N] $ are the quasideterminants of the particular solutions of QPII linear system (\ref{QPIIb}).\\ 
\textbf{Proof}\\
In order to prove this proposition, first we express the transformations (\ref{d3}) and (\ref{d4}) in terms of quasideterminants.
Now consider the transformation  (\ref{d3}) in the following form \[ \chi [1]= \gamma_{0} \Phi_{0} -\gamma_{1} \Phi_{1}(\gamma_{1}) \chi^{-1}_{1}(\gamma_{1}) \chi_{0}  \]or
\begin{equation}\label{qd1}
\chi [1]=
\begin{vmatrix}
 \chi_{1} & \chi_{0}\\
\gamma_{1} \Phi_{1} & {\boxed{\gamma_{0} \Phi_{0}}}
\end{vmatrix}
=\delta_{\chi}^{e}[1]
\end{equation}similarly we can do for the equation (\ref{d4})
\begin{equation}\label{qd2}
 \Phi [1]=
\begin{vmatrix}
 \Phi_{1} & \Phi_{0}\\
\gamma_{1} \chi_{1} & {\boxed{\gamma_{0} \chi_{0}}}
\end{vmatrix}=\delta_{\Phi}^{e}[1]
\end{equation} we have taken $\gamma=\gamma_{0}$, $\chi=\chi_{0}$ and $\Phi=\Phi_{0}$  in order to generalize the transformations in $N$th form.
Further,  we can represent the transformations $\chi [2]$ and $\Phi [2]$ by quasideterminants
 \[\chi [2]=
\begin{vmatrix}
 \chi_{2} & \chi_{1} & \chi_{0}\\
 \gamma_{2} \Phi_{2}& \gamma_{1} \Phi_{1} & \gamma_{0} \Phi_{0}\\
 \gamma^{2}_{2} \chi_{2} & \gamma^{2}_{1} \chi_{1}& {\boxed{\gamma^{2}_{0} \chi_{0}}}
\end{vmatrix}=\delta_{\chi}^{o}[2] \] and
 \[\Phi [2]=
\begin{vmatrix}
 \Phi_{2} & \Phi_{1} & \Phi_{0}\\
 \gamma_{2} \chi_{2} & \gamma_{1} \chi_{1} & \gamma_{0} \chi_{0}\\
 \gamma^{2}_{2} \Phi_{2} & \gamma^{2}_{1} \Phi_{1} & {\boxed{\gamma^{2}_{0} \Phi_{0}}}
\end{vmatrix}=\delta_{\Phi}^{o}[2]\] here superscripts $e$ and $o$  of $\delta$ represent the even and odd order quasideterminants.  The $N$th transformations 
for $\delta_{\chi}^{o}[N]$ and $\delta_{\Phi}^{o}[N]$ in terms of  quasideterminants are given below
  \[\delta_{\chi}^{o}[N]=
\begin{vmatrix}
 \chi_{N} & \chi_{N-1} & \cdots & \chi_{1} & \chi_{0}\\
\gamma_{N} \Phi_{N} & \gamma_{N-1} \Phi_{N-1} & \cdots & \gamma_{1}\chi_{1}& \gamma_{0} \Phi_{0}\\
 \vdots & \vdots & \vdots & \vdots & \vdots\\
 \gamma^{N-1}_{N} \Phi_{N} & \gamma^{N-1}_{N-1} \Phi_{N-1} & \cdots & \gamma^{N-1}_{1}\chi_{1} & \gamma^{N-1}_{0} \Phi_{0}\\
 \gamma^{N}_{N} \chi_{N} & \gamma^{N}_{N-1} \chi_{N-1} & \cdots & \gamma^{N}_{1} \chi_{1} & {\boxed{\gamma^{N}_{0} \chi_{0}}}
\end{vmatrix}\]and
  \[\delta_{\Phi}^{o}[N]=
\begin{vmatrix}
 \Phi_{N} & \Phi_{N-1} & \cdots& \Phi_{1} & \Phi_{0}\\
 \gamma_{N} \chi_{N} & \gamma_{N-1} \chi_{N-1} & \cdots & \gamma_{1}\chi_{1} & \gamma_{0} \chi_{0}\\
\vdots & \vdots & \vdots & \vdots & \vdots\\
 \gamma^{N-1}_{N} \chi_{N} & \gamma^{N-1}_{N-1} \chi_{N-1} & \cdots & \gamma^{N-1}_{1}\chi_{1} & \gamma^{N-1}_{0} \chi_{0}\\
 \gamma^{N}_{N} \Phi_{N} & \gamma^{N}_{N-1} \Phi_{N-1} & \cdots & \gamma^{N}_{1} \Phi_{1} & {\boxed{\gamma^{N}_{0} \Phi_{0}}}
\end{vmatrix} \]
here $N$ is to be taken as even.
In the same way we can write $N$th quasideterminant representations of $ \delta_{\chi}^{e}[N]$ and $ \delta_{\Phi}^{e}[N]$.
 \[ \delta_{\chi}^{e}[N ]=
\begin{vmatrix}
 \chi_{N} & \chi_{N-1} & \cdots & \chi_{1} & \chi_{0}\\
 \gamma_{N} \Phi_{N} & \gamma_{N-1} \Phi_{N-1} & \cdots& \gamma_{1}\chi_{1} & \gamma_{0} \Phi_{0}\\
 \vdots & \vdots&  \vdots & \vdots & \vdots\\
 \gamma^{N-1}_{N} \chi_{N} & \gamma^{N-1}_{N-1} \chi_{N-1} & \cdots& \gamma^{N-1}_{1}\chi_{1} & \gamma^{N-1}_{0} \chi_{0}\\
 \gamma^{N}_{N} \Phi_{N} & \gamma^{N}_{N-1} \Phi_{N-1} & \cdots & \gamma^{N}_{1} \Phi_{1} & {\boxed{\gamma^{N}_{0} \Phi_{0}}}
\end{vmatrix} \] 
and
\[\delta_{\Phi}^{e}[N]=
\begin{vmatrix}
 \Phi_{N} & \Phi_{N-1} & \cdots & \Phi_{1} & \Phi_{0}\\
 \gamma_{N} \chi_{N} & \gamma_{N-1} \chi_{N-1} & \cdots & \gamma_{1}\chi_{1} & \gamma_{0} \chi_{0}\\
 \vdots & \vdots & \vdots & \vdots & \vdots\\
 \gamma^{N-1}_{N} \Phi_{N}& \gamma^{N-1}_{N-1} \Phi_{N-1} & \cdots& \gamma^{N-1}_{1} \Phi_{1} & \gamma^{N-1}_{0} \Phi_{0}\\
 \gamma^{N}_{N} \chi_{N}& \gamma^{N}_{N-1} \chi_{N-1}& \cdots & \gamma^{N}_{1} \chi_{1} & {\boxed{\gamma^{N}_{0} \chi_{0}}}
\end{vmatrix}.\]
Similarly, we can derive the expression for $N$th soliton solution from equation (\ref{d11}) by applying the Darboux transformation iteratively, now consider
\[  u[1]=-4 \gamma \Omega_{1}^{\phi}[1]  \Omega_{1}^{\chi}[1]^{-1} +   \Omega_{1}^{\phi}[1]  \Omega_{1}^{\chi}[1]^{-1} u  \Omega_{1}^{\phi}[1]  \Omega_{1}^{\chi}[1]^{-1}\]
where  \[  \Omega_{1}^{\phi}[1]=\Phi_{1} \]
\[ \Omega_{1}^{\chi}[1] =\chi_{1}\]
this is one fold Darboux transformation. The two fold Darboux transformation is given by
 \begin{equation}\label{it2} 
  u[2]=  -4\gamma \phi[1]\chi^{-1}[1] + \phi[1]\chi^{-1}[1]u[1]\phi[1]\chi^{-1}[1].
 \end{equation} 
 We may rewrite the equation (\ref{qd1}) and equation (\ref{qd2}) in the following forms
\[\chi [1]=
\begin{vmatrix}
 \chi_{1} & \chi_{0}\\
\gamma_{1} \Phi_{1} & {\boxed{\gamma_{0} \Phi_{0}}}
\end{vmatrix}
= \Omega_{2}^{\chi}[2]
\]

 \[\Phi [1]=
\begin{vmatrix}
 \Phi_{1} & \Phi_{0}\\
\gamma_{1} \chi_{1} & {\boxed{\gamma_{0} \chi_{0}}}
\end{vmatrix}= \Omega_{2}^{\phi}[2].\]
and equation (\ref{it2}) may be written as
\[u [2]=  \varTheta^{(-)}_{2} (  -4\gamma +  u ) \varTheta^{(+)}_{2}  \]
where
\[ \varTheta^{(-)}_{2} =  \Omega_{2}^{\phi}[2]  \Omega_{2}^{\chi}[2]^{-1}  \Omega_{1}^{\phi}[1]  \Omega_{1}^{\chi}[1]^{-1} \]
\[  \varTheta^{(+)}_{2} =   \Omega_{1}^{\phi}[1] \Omega_{1}^{\chi}[1]^{-1} \Omega_{2}^{\phi}[2]  \Omega_{2}^{\chi}[2]^{-1} \]
In the same way, we can derive the expression for three fold Darboux transformation
\[u[3]= \varTheta^{(-)}_{3}  (  -4\gamma +  u )  \varTheta^{(+)}_{3} \]
where 
\[ \varTheta^{(-)}_{3} =  \Omega_{3}^{\phi}[3]  \Omega_{3}^{\chi}[3]^{-1}  \Omega_{2}^{\phi}[2]  \Omega_{2}^{\chi}[2]^{-1} \Omega_{1}^{\phi}[1] \Omega_{1}^{\chi}[1]^{-1} \]
\[  \varTheta^{(+)}_{3} =   \Omega_{1}^{\phi}[1] \Omega_{1}^{\chi}[1]^{-1} \Omega_{2}^{\phi}[2]  \Omega_{2}^{\chi}[2]^{-1} \Omega_{3}^{\phi}[3]  \Omega_{3}^{\chi}[3]^{-1}\]
with following settings
 \[\chi [2]=
\begin{vmatrix}
 \chi_{2} & \chi_{1} & \chi_{0}\\
 \gamma_{2} \Phi_{2}& \gamma_{1} \Phi_{1} & \gamma_{0} \Phi_{0}\\
 \gamma^{2}_{2} \chi_{2} & \gamma^{2}_{1} \chi_{1}& {\boxed{\gamma^{2}_{0} \chi_{0}}}
\end{vmatrix}=  \Omega_{3}^{\chi}[3]\] and
 \[\Phi [2]=
\begin{vmatrix}
 \Phi_{2} & \Phi_{1} & \Phi_{0}\\
 \gamma_{2} \chi_{2} & \gamma_{1} \chi_{1} & \gamma_{0} \chi_{0}\\
 \gamma^{2}_{2} \Phi_{2} & \gamma^{2}_{1} \Phi_{1} & {\boxed{\gamma^{2}_{0} \Phi_{0}}}
\end{vmatrix}= \Omega_{3}^{\phi}[3].\] 

Finally, by applying the  transformation iteratively we can construct the $N$-fold Darboux transformation
\[u[N]= \Omega_{N}^{\phi}[N]  \Omega_{N}^{\chi}[N]^{-1}  \Omega_{N-1}^{\phi}[N-1]  \Omega_{N-1}^{\chi}[N-1]^{-1}...  \Omega_{2}^{\phi}[2]  \Omega_{2}^{\chi}[2]^{-1}  \Omega_{1}^{\phi}[1]  \Omega_{1}^{\chi}[1]^{-1} \times \]
\[ \times (  -4\gamma +  u )   \Omega_{1}^{\phi}[1]  \Omega_{1}^{\chi}[1]^{-1}  \Omega_{2}^{\phi}[2]  \Omega_{2}^{\chi}[2]^{-1}...  \Omega_{N-1}^{\phi}[N-1]  \Omega_{N-1}^{\chi}[N-1]^{-1}\gamma_{N}^{\phi}[N] \gamma_{N}^{\chi}[N]^{-1}\]
 by considering the following substitution 
\[\Theta_{N}[N]= \Omega_{N}^{\phi}[N]  \Omega_{N}^{\chi}[N]^{-1}\]
in above expression, we get
\[u[N]=\Theta_{N}[N]\Theta_{N-1}[N-1]...\Theta_{2}[2]\Theta_{1}[1] (  -4\gamma +  u )  \Theta_{1}[1]\Theta_{2}[2]...\Theta_{N-1}[N-1]\Theta_{N}[N]\]
or
\begin{equation}\label{NCPDTS} 
 u[N]=\Pi^{N-1}_{k=0} \Theta_{N-k}[N-k]  (  -4\gamma +  u )  \Pi^{0}_{j=N-1}\Theta_{N-j}[N-j]
\end{equation} 
here we present only the  $N$th expression for odd order quasideterminants  $  \Omega_{N}^{\phi}[N]$ and  $  \Omega_{N}^{\chi}[N]$ 
 \[  \Omega_{2N+1}^{\phi}[2N+1]=
\begin{vmatrix}
 \Phi_{2N} & \Phi_{2N-1} & \cdots& \Phi_{1} & \Phi_{0}\\
 \gamma_{2N} \chi_{2N} & \gamma_{2N-1} \chi_{2N-1} & \cdots & \gamma_{1}\chi_{1} & \gamma_{0} \chi_{0}\\
\vdots & \vdots & \vdots & \vdots & \vdots\\
 \gamma^{2N-1}_{2N} \chi_{2N} & \gamma^{2N-1}_{2N-1} \chi_{2N-1} & \cdots & \gamma^{2N-1}_{1}\chi_{1} & \gamma^{2N-1}_{0} \chi_{0}\\
 \gamma^{2N}_{2N} \Phi_{2N} & \gamma^{2N}_{2N-1} \Phi_{2N-1} & \cdots & \gamma^{2N}_{1} \Phi_{1} & {\boxed{\gamma^{2N}_{0} \Phi_{0}}}
\end{vmatrix} \]
and

 \[  \Omega_{2N+1}^{\chi}[2N+1]=
\begin{vmatrix}
 \chi_{2N} & \chi_{2N-1} & \cdots & \chi_{1} & \chi_{0}\\
\gamma_{2N} \Phi_{2N} & \gamma_{2N-1} \Phi_{2N-1} & \cdots & \gamma_{1}\Phi_{1}& \gamma_{0} \Phi_{0}\\
 \vdots & \vdots &  \vdots & \vdots & \vdots\\
 \gamma^{2N-1}_{2N} \Phi_{2N} & \gamma^{2N-1}_{2N-1} \Phi_{2N-1} & \cdots & \gamma^{2N-1}_{1} \Phi_{1} & \gamma^{2N-1}_{0} \Phi_{0}\\
 \gamma^{2N}_{2N} \chi_{2N} & \gamma^{2N}_{2N-1} \chi_{2N-1} & \cdots & \gamma^{2N}_{1} \chi_{1} & {\boxed{\gamma^{2N}_{0} \chi_{0}}}
\end{vmatrix} .\]
\section{Conclusion}
In this paper, I have presented the zero curvature  representation of non-abelian quantum PII equation and also its Darboux transformation which are further applied to construct the multi-solution expression to that equation in terms quasideterminants.
 I  have also derived  Riccati equation associated to non-abelian quantum PII equation from its linear system by using the method of Konno and Wadati  \cite{KKMW}.  Symmetrically it is quite interesting  to construct zero curvature representations for quantum Painlev\'e equations PIV, PV by using the procedure presented in this article to derive non-abelian quantum analogues of  these equations involving the symmetric forms of these systems  given in \cite{NH}.  
 \section{Acknowledgement}
I would like to thank V. Roubtsov and V. Retakh  for their  discussions to me during my  Ph. D. research work.
 My special thanks to the University of the Punjab, Pakistan, on funding me for my Ph.D. project in France. I am also thankful to LAREMA, Universit\'e d'Angers and ANR "DIADEMS", France on providing me facilities during my Ph.D. research work.


\noindent
\begin{thebibliography}{99}

\bibitem{reb}E. Br\'ezin and V. A. Kazakov, Exactly stable field theories of closed strings, Phys. Letts. B 236 (1990) 144-150.
\bibitem{rdg}D. Gross and A. Migdal, A nonperturbative treatment of two dimen- sional quantum gravity, Nucl. Phys. B 340 (1990) 33-365.
\bibitem{rasf} A. S. Fokas and S. Tanveer, A Hele-Shaw problem and second Painlev\'e transcendent?, Math. Proc. Comb. Phil. Soc. 124 (1998) 169-191.
\bibitem{rebm}E. Barouch, B. M. McCoy, C. A. Tracy and T .T. Wu, Zero field susceptibility of the two-dimensional Ising model near Tc, Phys.Rev.Lett.31(1973)1409-1411.
\bibitem{rmj} M. Jimbo, T. Miwa and Y. Mori, ?M.Sato,Density matrix of impenetra- ble bose gas and the fifth Painlev?e transcendent,Physica D1 (1980)80- 158.
\bibitem{ras}A. S .Fokas, A. R. Its, A. A. Kapaev and V.Y. Novokshenov, Painlev\'e transcendents: The Riemann-Hilbert approach, Mathematical surveys and monographs, ISSN 0076-5376; v.128.
\bibitem{rri}R. Iyer, C. V. Johnson and J. S. Pennington,String Theory and Water Waves, SLAC-PUB-15670.
\bibitem{r5}A. N. W. Hone, Painlev\'e Test, Singularity Structure and Integrability, Lect. Notes Phys. \textbf{ 767} , 245-277 (2009).
\bibitem{r11}K. Okamoto, in: R. Conte (Ed.), The Painlev\'e property, One century later, CRM Series in Mathematical Physics, Springer, Berlin, (1999) 735-787.
\bibitem{r12}S. P. Balandin, V. V. Sokolov, On the Painlev\'e test for non-abelian equations, Physics letters, \textbf{A246} (1998) 267-272.
\bibitem{r4}N. Joshi, The second Painlev\'e hierarchy and the stationarty KdV hierarchy, Publ. RIMS, Kyoto Univ. \textbf{ 40}(2004) 1039-1061.
\bibitem{MA}N. Joshi, M. Mazzocco, Existence and uniqueness of tri-tronqu\'ee solutions of the second Painlev\'e hierarchy, Nonlinearity \textbf{ 16}(2003) 427--439. 
\bibitem{NH} H. Nagoya, Quantum Painlev\'e systems of type Al , Internat. J. Math. 15 (2004), 1007–1031,math.QA/0402281.
\bibitem{NGR} H. Nagoya, B. Grammaticos and A. Ramani, Quantum Painlev\'e equations: from Continuous to discrete, SIGMA \textbf{ 4} (2008), no. 051, 9 pages.
\bibitem{NY} M. Noumi and Y. Yamada, Higher order Painlev\'e equations of type Al , Funkcial. Ekvac. \textbf{ 41} (1998) 483?503, math.QA/9808003.
\bibitem{7}V. Retakh, V. Rubtsov, Noncommutative Toda chain , Hankel quasideterminants and Painlev\'e II equation, J. Phys. A: Math. Theor. \textbf{ 43}   (2010)  505204 (13pp)

\bibitem{MIRFAN} M. Irfan, Lax pair representation and Darboux transformation of noncommutative Painlev\'e's second equation,Journal of Geometry and Physics, \textbf{ 62} (2012) 15751582.
\bibitem{IM}  Irfan Mahmood, Quasideterminant solutions of NC Painlev\'e II equation with the Toda solution at $n=1$ as a seed solution in its Darboux transformation,  Joural of geometry and physics 95 (2015) 127-136. 
\bibitem{PL}P. Lax. Integrals of nonlinear equations of evolution and solitary waves, Commun. Pure Appl. Math., XXI: 467490, 1968.
\bibitem{r22}A. Dimakis, F. M. Hoissen, Noncommutative Korteweg-de Vries equation, Preprint hep-th/0007074, (2000).
\bibitem{r19} B. A. Kupershmidt, Noncommutative integrable systems, in nonlinear evolution equations and dynamical systems, NEEDS 1994, V. Makhankov et al ed-s,  World Scientific (1995) 84-101.
\bibitem{r16}L. D. Paniak, Exact noncommutative KP and KdV multi-solitons, hep-th/0105185.
\bibitem{r13}M. Hamanaka, K. Toda, Towards noncommutative integrable systems, Phys. Lett. \textbf{ A316}(2003) 77.
\bibitem{r24}M. Hamanaka, K. Toda , Noncommutative Burgers equation, J. Phys. hep-th/0301213, \textbf{ A36}(2003) 11981.
\bibitem{r23}M. Legar\'e, Noncommutative generalized NS and super matrix KdV systems from a noncommutative version of (anti-) selfdual Yang-Mills equations, Preprint hep-th/0012077, 2000.
\bibitem{e2}A. Dimakis, F. M. Hoissen, With a Cole-Hopf transformation to solution of noncommutative KP hierarchy in terms of Wronski martices, J. Phys. \textbf{ A40}(2007)F32.
\bibitem{4}A. Dimakis, F. M. Hoissen, The Korteweg-de Vries equation on a noncommutative space-time, Phys. Lett. \textbf{ A278}(2000) 139-145.
\bibitem{3}I. C. Camero, M. Moriconi, Noncommutative integrable field theories in 2d, Nucl. Phys. \textbf{ B673}(2003) 437-454.
\bibitem{1}S. Carillo, C. Schieblod, Noncommutative Korteweg-de Vries and modified Korteweg-de Vries hierarchies via recursion methods, J. Math. Phys. \textbf{ 50}(2009) 073510.
\bibitem{KKMW} K. Konno and M. Wadati,  Simple Derivation of B\"acklund transformation from Riccati Form of inverse method, Progress of Theoretical Physics, Vol. 53, No. 6, June (1975).
\bibitem{VM} V. B. Matveev, M. A. Salle, Darboux Transformations and Soliton, Berlin: Springer,1991.
\bibitem{GelRet} I. Gelfand, S. Gelfand V. Retakh, R. L. Wilson, Quasideterminants , Advances in Mathematics \textbf{ 193}  (2005) 56-141.
\bibitem{r26}C. R. Gilson,  J. J. C. Nimmo, On a direct approach to quasideterminant solutions of a noncommutative KP equation, J. Phys.






 \end{thebibliography}

\end{document}